\documentclass[%
reprint,
nofootinbib,
 amsmath,amssymb,
 aps,
]
{revtex4-1}
\usepackage{epsfig,epsf,rotating,wrapfig,tabularx}
\usepackage{dcolumn}
\usepackage{bm}
\usepackage{graphicx}

\begin{document}

\preprint{APS/123-QED}

\title{Secondary hadron distributions in two component model}

\author{\firstname{A.~A.}~\surname{Bylinkin}}
 \email{alexandr.bylinkin@cern.ch}
\affiliation{%
 Institute for Theoretical and Experimental
Physics, ITEP, Moscow, Russia
}%
\author{\firstname{M.~G.}~\surname{Ryskin}}
 \email{ryskin@thd.pnpi.spb.ru}
\affiliation{%
Petersburg Nuclear Physics Institute, NRC Kurchatov Institute, Gatchina,
St. Petersburg, 188300, Russia.
}%


\begin{abstract}
Inclusive charged hadron cross sections, $d\sigma/d\eta$, and the mean transverse momenta, $<p_T>$, are considered within the two component model, which combines the power-like and the exponential terms in $p_T$. The observed dependences of the spectra shape on energy and the event multiplicity qualitatively agree with that expected from the Regge theory with the perturbative QCD pomeron. Finally, the dependences observed are used to make predictions on the mean transverse momenta, $<p_T>$ as function of multiplicity at LHC-energies, which are tested on available experimental data.
\end{abstract}

\pacs{Valid PACS appear here}
\maketitle


At high energies, $\sqrt s$, the hadron-hadron interaction and the multiparticle production are usually considered in terms of the pomeron exchange. Besides the single pomeron exchange there are more complicated contributions described by the multi-pomeron diagrams in the framework of the Reggeon Field Theory (RFT)\cite{RFT}. This multi-pomeron terms account for the absorptive corrections and rescattering effects.

In perturbative QCD  (pQCD) the pomeron exchange amplitude is given by the set of the "ladder-like" diagrams build up of the (reggeized) gluons. These diagrams sum up all the contributions where the small value of the QCD coupling constant $\alpha_s$ is compensated by the large value of $\ln s$~\cite{IL}. In the Leading Log approximation the intercept of this BFKL pomeron $\alpha_P(0)=1+\Delta$ turns out to be rather 
large~\cite{BFKL}:
\begin{equation} 
\Delta=\frac{\alpha_s N_c}{\pi}\cdot4\ln 2\ .
\end{equation}
Numerically it leads to $\Delta>0.5$. Accounting for the 'next-to-leading-Log' (NLL) corrections we get a lower, but still rather large intercept. The resummation of the NLL contributions gives $\Delta\sim 0.2-0.3$~\cite{NNL}.

On the other hand, the high energy cross-sections grow much slowly, like $\sigma_{tot}\propto s^{\Delta_{eff}}$ with $\Delta_{eff}\sim 0.1$~\cite{DL}.  This fact is explained by large absorptive corrections caused by the multi-pomeron diagrams. Due to so large absorptive (multi-pomeron) effects it is not easy to study the properties of the {\em individual} pomeron experimentally.

\begin{figure}[h]
\includegraphics[width =8cm]{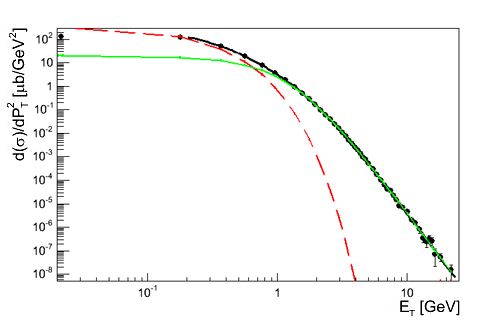}
\caption{\label{fig.0} Charged particle spectrum~\cite{UA1} fitted to the function~(\ref{eq:exppl}): the red (dashed) curve shows the exponential term and the green (solid) one stands for the power-like term.}
\end{figure}

Therefore, it is interesting to see the energy dependence of the inclusive cross sections (and the mean transverse momenta, $<p_T>$, of secondaries) obtained from the fit where their spectra are described using the {\em two component} ansatz~\cite{OUR1} including two qualitatively different contributions: with a power and an exponential $p_T$ behavior:
\begin{equation}
\label{eq:exppl}
\frac{d\sigma}{p_T d p_T} = A_e\exp {(-E_{Tkin}/T_e)} +
\frac{A}{(1+\frac{p_T^2}{T^{2}\cdot n})^n},
\end{equation}
where  $E_{Tkin} = \sqrt{p_T^2 + M^2} - M$
with M equal to the produced hadron mass. $A_e, A, T_e, T, n$ are the free parameters to be determined by fit to the data.
The power-like component is mainly originated from a relatively large $p_T$ domain, that is from the mini-jet fragmentation, while the exponential part accounts for some "thermalization" caused by the final state rescattering in the parton cloud and the "hadron gas" formed by the secondaries (Fig.~\ref{fig.0}).\\

Recall that according to the AGK cutting rules~\cite{AGK} the corrections caused by the multi-pomeron graphs are almost absent in the single particle inclusive cross sections $d\sigma(a+b\to c+X)/d^3p$. Indeed, in this case there are no contribution from the diagrams where an additional pomeron crosses the rapidity
of detected particle $c$. This means that here we have no corrections from the non-enhanced, eikonal-like diagrams. Within the eikonal (or multi-channel eikonal) models such inclusive cross section is described just by the {\em one} pomeron exchange. Moreover, thanks to the AGK rules, an important part of the 'enhanced'
absorptive corrections (which account for the rescattering and the interactions 
between the intermediate particles in the pomeron ladder and are "enhanced" by the multiplicity of these intermediate particles) is canceled as well. The remaining  enhanced diagrams, corresponding to the interactions between the intermediate particles in only one hemisphere (between the hadrons $a$ and $c$ or $b$ and $c$) 
are suppressed for the power-like part of the spectra due to a relatively large $q_t$ of the original mini-jet which acts as a source of this power-like component.

 Thus, the behavior of the power-like part of single particle inclusive cross 
 sections provides a most direct information about the 'bare' pomeron properties.
In particular, we expect that the particle density $d\sigma^{power}/d\eta$ should increase with energy as 
\begin{equation}
\frac{d\sigma^{power}}{d\eta}~\propto ~ s^{\Delta_P}\ ,
\end{equation}
where $\alpha_P(0)=1+\Delta_P$ is the true intercept of the initial (bare) pomeron.\\

\begin{figure}[h]
\includegraphics[width =8cm]{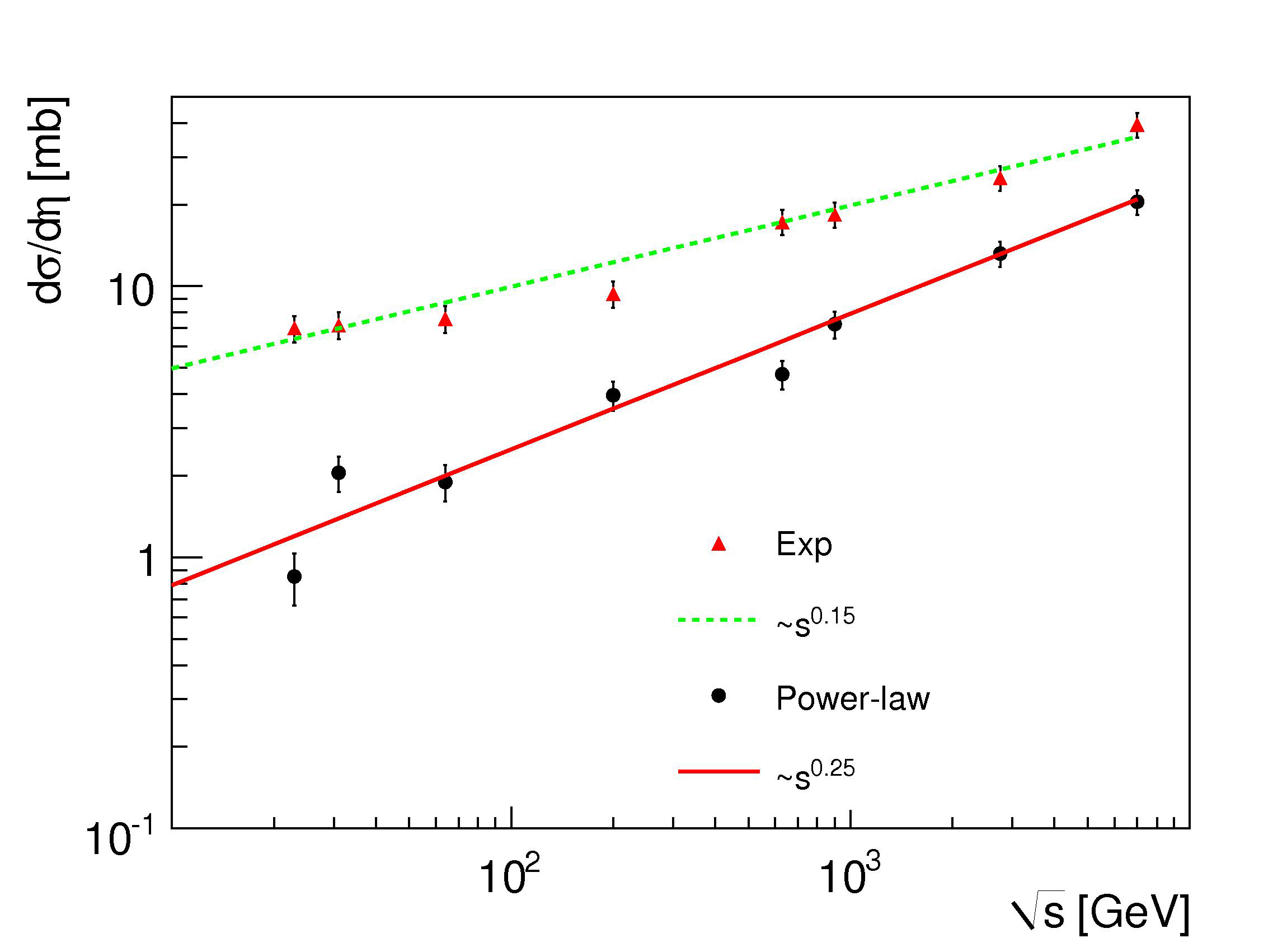}
\caption{\label{fig.1} Particle densities $d\sigma/d\eta$ as function of c.m.s energy $\sqrt{s}$ in $pp$-collisions~\cite{ISR, PHENIX, UA1, ALICE} calculated for power-like and exponential contributions of~(\ref{eq:exppl}) separately. }
\end{figure}

This effect can be studied by fitting available experimental data  on charged hadron production in $pp$-collisions from ISR to LHC energies~\cite{ISR, PHENIX, UA1, ALICE} by the parameterisation introduced~(\ref{eq:exppl}) and integrating power-like and exponential contributions separately over $p_T^2$. As it is seen in Fig.~\ref{fig.1} for the power-like part of the spectra we observe $\Delta\simeq 0.25$ - close to the value expected for the pQCD (BFKL) pomeron after the resummation of the NLL corrections. The value of $\Delta$ coming from the 'exponential' component is lower ($\simeq 0.15$) since it is strongly affected by absorptive corrections.\\

\begin{figure}[h]
\includegraphics[width =8cm]{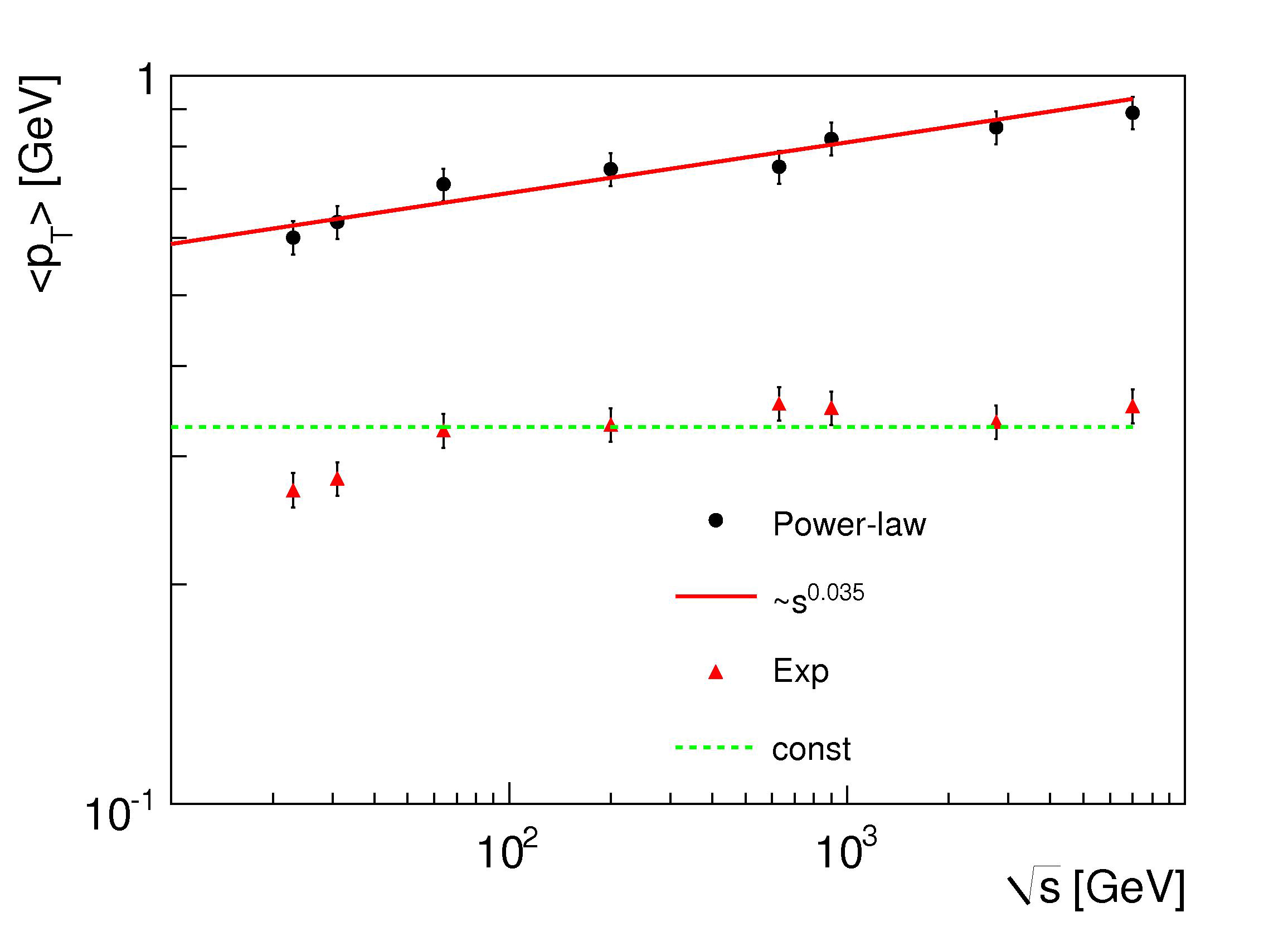}
\caption{\label{fig.2} Mean transverse momenta $<p_T>$ as function of c.m.s energy $\sqrt{s}$ in $pp$-collisions ~\cite{ISR, PHENIX, UA1, ALICE} calculated for power-like and exponential contributions of~(\ref{eq:exppl}) separately. }
\end{figure}

Let us now discuss the behavior of the mean transverse momenta, $<p_T>$,  of secondaries corresponding to each component. The energy dependence of $<p_T>$ is presented in Fig.~\ref{fig.2}. While the value of $<p_T>$ in the 'exponential' part is almost constant (except of the low energy domain where some growth of $<p_T>$ may be caused just by the kinematical reasons) the $<p_T>$ corresponding to the power-like part increases as $<p_T>\propto (\sqrt s)^{0.07}$. This may be explained by the growth of the typical transverse momenta of mini-jets, which hadronization is responsible for this part of the spectra.
Recall that the perturbative QCD is the theory with dimensionless coupling, $\alpha_s$. Therefore, the distribution over the transverse momentum in the next cell of the pomeron ladder is driven by the integral which has the logarithmic form like
\begin{equation}
\int\frac{\alpha_s(k_t^2)d^2k_t}{k^2_t+k^{'2}_t}\ ,
\end{equation}
where the value of $k_t$ is determined by  $k'_t$ of the t-channel gluon in the neighbouring cell. Due to the up/down symmetry at each evolution step the $k_t$ of gluon may become few times larger or smaller with equal probabilities. This way we obtain the famous BFKL-diffusion in $\ln k_t$ space~\cite{L}. Accounting for the running $\alpha_s(k^2)$ shifts the 'diffusion' to the direction of a lower $k_t$. On the other hand, the low $k_t$ gluons have a large probability to be absorbed, $\sigma^{abs}\sim 1/k^2_t$. The last effect turns out to be stronger. Thus, finally, the absorptive corrections 
(described by the enhanced diagrams where the pomeron-loop is completely concentrated between one of the incoming hadrons, say, $a$ and the observed particle $c$)  push the BFKL-diffusion to enlarge the $k_t(s)$. It was shown that asymptotically at very large $s\to\infty$ the mini-jet transverse momentum increases as
\begin{equation}
<k_t>\sim \Lambda_{QCD}\exp(1.26\sqrt{\ln s})
\end{equation}
(see for example~\cite{LR}). This growth of the mini-jet $k_t$ leads to the increase of the mean $<p_t>$ of secondaries described by the power-like 
component of the spectra.\\

Next we consider the dependence of $<p_T>$ on the event multiplicity $N_{ch}$.
Within the Regge theory the higher multiplicity events are described by the diagrams with a larger number, $n$, of 'cut' pomerons ($N_{ch}\propto n$). 
In the old Regge theory, where the values of $p_T$ were limited, we have to expect the same $p_T$ distribution of the particles in the central (plato) region independently on the number of cut pomerons in the event. Therefore -- $<p_T>=const(N_{ch})$, since each/any Pomeron produces the same $p_T$-spectra, while $dN/d\eta\propto N_{ch}$.

\begin{figure}[h]
\includegraphics[width =8cm]{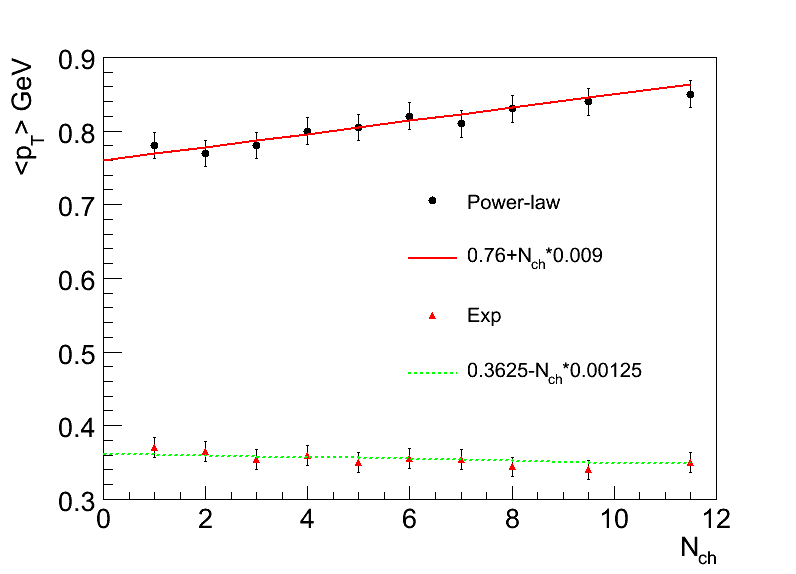}
\caption{\label{fig.3} Mean transverse momenta $<p_T>$ of power-like and exponential contributions of~(\ref{eq:exppl}) as function of multiplicity $N_{ch}$ calculated from the  fits to the STAR-data~\cite{STAR}. }
\end{figure}

Accounting for mini-jet (pQCD) contribution the $<p_T>$ should increase  with $N_{ch}$ since another way to enlarge multiplicity is to produce mini-jets with
a larger $E_T$ (and thus a larger multiplicity in jet-fragmentation). Indeed, in Fig.~\ref{fig.3} 
we see that for the power-like component the value of $<p_T>$ slightly increases with $N_{ch}$ while for the exponential part it a bit decreases (due to some 'cooling' during the expansion of a larger multiplicity system).

\begin{figure}[h]
\includegraphics[width =8cm]{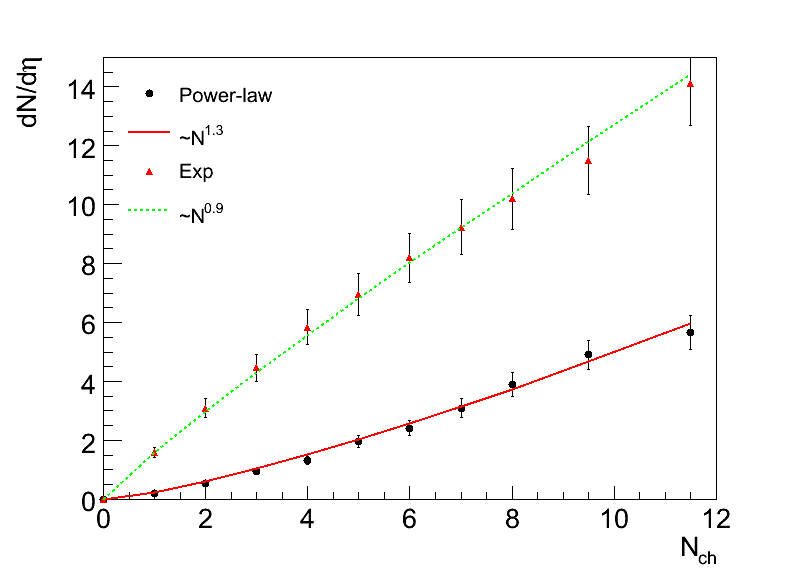}
\caption{\label{fig.4} Charged particle densities $dN/d\eta$ of power-like and exponential contributions of~(\ref{eq:exppl}) as function of multiplicity $N_{ch}$ calculated from the  fits to the STAR-data~\cite{STAR}. }
\end{figure}

In addition, the dependence of charged particle densities $dN/d\eta$ on the event multiplicity $N_{ch}$ can be considered. According to the AGK rules~\cite{AGK} the charge multiplicity in hadronic interactions is proportional  to the number of pomerons involved. This behavior can also be studied by fitting the STAR-data~\cite{STAR} represented by charged particle spectra for events with different visible charged multiplicity using the ansatz introduced~(\ref{eq:exppl}).
Figure~\ref{fig.4} shows the increase of charged particle densities $dN/d\eta$ from the power-like and exponential contributions of ~(\ref{eq:exppl}) as function of multiplicity $N_{ch}$. One can notice that the contribution from the power-like
component (mini-jets) grows faster than that from the 'exponential' one ($N_{ch}^{1.3}$ vs $N_{ch}^{0.9}$). This fact confirms that partly the increase of $N_{ch}$ is due to a larger $E_T$ of mini-jets produced.

Finally, the observed dependences (Figs.~\ref{fig.1}-\ref{fig.4}) can be used to make predictions on the mean transverse momenta, $<p_T>$, as function of multiplicity $N_{ch}$ at higher energies and tested on the available data from LHC~\cite{ALICEN}. The values of $<p_T>$ and $dN/d\eta$ as function of $N_{ch}$ are  parametrized from the STAR-data~\cite{STAR} as outlined in Figs.~\ref{fig.3} and \ref{fig.4}:

\begin{equation}
<p_T^{power}>(N_{ch}) = 0.76+N_{ch}^{0.009},
\end{equation}
\begin{equation}
<p_T^{exp}>(N_{ch}) = 0.3625-N_{ch}^{0.00125}.
\end{equation}
\begin{equation}
(\frac{dN}{d\eta})^{power}(N_{ch}) = 0.3\cdot N_{ch}^{1.3},
\end{equation}
\begin{equation}
(\frac{dN}{d\eta})^{exp}(N_{ch}) = 1.5\cdot N_{ch}^{0.9}.
\end{equation}
 Then these dependences are extrapolated to higher energies from those observed in Figs.~\ref{fig.1} and \ref{fig.2}:

\begin{equation}
<p_T^{power}>(N_{ch}, s) = <p_T^{power}>(N_{ch})\cdot(\frac{s}{s_0})^{0.07},
\end{equation}
\begin{equation}
<p_T^{exp}>(N_{ch}, s) = <p_T^{exp}>(N_{ch})\cdot(\frac{s}{s_0})^{0.03},
\end{equation} 
\begin{equation}
(\frac{dN}{d\eta})^{power}(N_{ch}, s) = (\frac{dN}{d\eta})^{power}(N_{ch})\cdot(\frac{s}{s_0})^{0.25},
\end{equation}
\begin{equation}
(\frac{dN}{d\eta})^{exp}(N_{ch}, s) = (\frac{dN}{d\eta})^{exp}(N_{ch})\cdot(\frac{s}{s_0})^{0.15},
\end{equation}
where $\sqrt{s_0} = 200$ Gev (STAR c.m.s. energy) and $\sqrt{s}$ - c.m.s. energy at LHC\footnote{Weak energy dependence $<p_T^{exp}> \propto s^{0.03}$ gives a better description of the data shown in Fig.~\ref{fig.2}.}. 

Finally, the resulting $<p_T>$ for higher energy is calculated by the formula:
\begin{widetext}
\begin{equation}
<p_T>(N_{ch}, s) = \frac{(\frac{dN}{d\eta})^{power}(N_{ch}, s)\cdot <p_T^{power}>(N_{ch}, s) + (\frac{dN}{d\eta})^{exp}(N_{ch}, s)\cdot <p_T^{exp}>(N_{ch}, s)}{(\frac{dN}{d\eta})^{power}(N_{ch}, s)+(\frac{dN}{d\eta})^{exp}(N_{ch}, s)}.
\end{equation}
\end{widetext}

\begin{figure}[h]
\includegraphics[width =8cm]{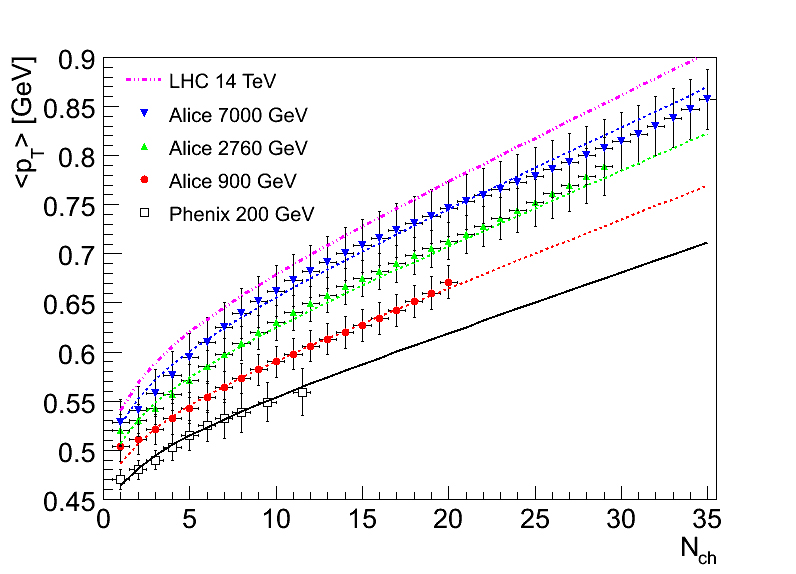}
\caption{\label{fig.5} Predictions (dashed lines) on the mean transverse momenta $<p_T>$ as function of multiplicity $N_{ch}$ calculated from the  fits (Figs.~\ref{fig.3} and \ref{fig.4}) to the STAR-data~\cite{STAR} (open points, solid line) for LHC-energies shown together with the ALICE-data~\cite{ALICEN} (full points). }
\end{figure}

 The results of this procedure are shown in Fig.~\ref{fig.5} for LHC-energies together with the data measured by the ALICE-collaboration~\cite{ALICEN}\footnote{Additional correction on $<p_T>$-predictions accounting the phase space $0.15 < p_T < 10$ GeV of the ALICE measurements~\cite{ALICEN} is introduced.} in $pp$-colllisions at $900$, $2760$ and $7000$ GeV, respectively.  STAR data (at $\sqrt{s} = 200$ GeV) together with this parameterization is shown as well.
One can notice a rather good agreement between the prediction and the experimental data measured by the ALICE collaboration at different energies. Thus, the prediction for further measurements at LHC at $14$ TeV can be made (Fig.~\ref{fig.5}).

In conclusion, inclusive charged hadron cross-sections, $d\sigma/d\eta$, and the mean transverse momenta, $<p_T>$, have been considered within the two component model as a function of c.m.s. energy $\sqrt{s}$ and multiplicity $N_{ch}$. The observed dependences have been discussed and shown to qualitatively agree with that expected from the Regge theory with the perturbative QCD pomeron. Finally, they have been used to make predictions on the mean transverse momenta, $<p_T>$ as function of multiplicity at LHC-energies. These predictions have been tested on available
experimental data and a good agreement has been observed. The prediction
for further LHC measurements at $14$ TeV has been made.\\


\begin{acknowledgements}
The authors thank Professor Andrey Rostovtsev for stimulating discussions. \\
\end{acknowledgements}

\end{document}